\documentclass{article}%
\usepackage{amsmath}
\usepackage{amsfonts}
\usepackage{amssymb}
\usepackage{graphicx}%
\begin{document}
\title{On E0 Transitions in even-even nuclei}
\author {Vladimir P. $Garistov^{1}$, A.A. $Solnyshkin^{2}$ , I.
$Adam^{2}$,\\
 O.K. $Egorov^{3}$, A. $Islamov^{3}$, V.I. $Silaev^{3}$, D.D.$Bogachenko ^{3}$}

\date{}
\maketitle

 {\small  {\center \hspace{10mm}Institute for Nuclear Research and
Nuclear Energy, Sofia, $Bulgaria^1$}

  {\center \hspace{15mm}Joint Institute for Nuclear Research, Dubna,$
Russia^2$}

  {\center \hspace{10mm}Institute for Theoretical and Experimental Physics,Moscow,
$Russia^3$}}
\begin{abstract}

\hspace{0.3in}The reanimation of the investigations dedicated to
$0^{+}$ states energies and $E0$ transitions between them is
provoked by new and more precise experimental technics that not
only made revision of the previous data  but also gave a
possibility to obtain a great amount of new $0^{+}$ states
energies and conversion electrons data. We suggest one
phenomenological model for estimation of the $E0$ transition
nuclear matrix elements. Recently theoretical
calculations$\cite{Vova}$ predicted existence of a $0^{+}$ state
with energy 0.68 MeV in $^{160}Dy$ nucleus. Powerful enough
arguments in favor of existence of 681.3 keV state in $^{160}Dy$
nucleus   are presented.
\end{abstract}

\section{Introduction}

\hspace{0.3in}Nature of low lying 0$^{+}$ states bands in deformed
nuclei remains a mystery under debate. The improvements in
technology have remedied the situation by enabling spectroscopy,
reactions, and life-time measurements of a large number of
K$^{\pi}$=0$^{+}$ bands that were previously inaccessible in
nuclei \cite{Aprahamian}. Some authors point out importance to
study anharmonic effects in microscopic way in deformed nuclei
\cite{Brac}, quadrupole and pairing vibrational modes in
conversional electrons and internal pair decay \cite{Toiv}
 , or exact diagonalization in the restricted space
of collective phonons of different types \cite{Kantele}.

\hspace{0.3in}The energies and electromagnetic decay properties of
the excited $0^{+}$ states are important in determining the
applicability and test of the models - Shell model,
Claster-vibrational model, Quasi-particle - phonon model, a
deformed configuration mixing shell model, Interacting boson
approximation, pairing quadrupole correlations, O(6)limit of
$IBA$.

    There are some but poor enough calculations
[ \cite{Toiv},
 \cite{Kumpulainen},\cite{Passoja}, \cite{J. Kantele}] devoted to estimation of $E0$ nuclear matrix elements
$\rho^{2}$ between different $0^{+}$ states in the same nucleus.
For instance in \cite{J. Kantele} we find that
$\rho^{2}(0^{+}_{2}\rightarrow0^{+}_{1})$ is very small in compare
with $\rho^{2}(0^{+}_{3}\rightarrow0^{+}_{1})$ that indicates the
more collective nature of the $0^{+}_{2}$ state. It should be very
important to determine the half-lives of the $0^{+}$ states that
would allow more definite conclusions on the structures of the
excited $0^{+}$ states  \cite{Passoja} . Often the first excited
$0^{+}$ state in nuclei is considered as less collective than the
next by increasing energy states. In some even-even nuclei the
first by energy excited state is not necessarily the lowest state
by degree of collectivity \cite{J. Kantele}. For instance observed
in $^{158}Gd\,\,$ $0^{+}$ state with the excitation energy 0,2548
MeV ($n=20$) \cite {Vova} is also much more collective than the
$0^{+}$ state with energy 0,5811 ($n=1$) MeV.

\section {E0 transition matrix elements}

\hspace{0.3in}The separation of the E0 conversion probability into
electronic and nuclear factors is not as well defined as for the
conversion of higher multipoles, nor is the electronic factor,$
\Omega$, completely independent of nuclear properties. Physically,
the monopole transition interaction vanishes except while the
electron is within the nuclear charge distribution, and hence it
is the electron wave functions within the nucleus which enter into
the calculation of $\Omega$. These in turn, depend on the average
static nuclear charge distribution. In this paper we will estimate
the transition probability without mutual influence of the
electronic and nuclear wave functions (that really may occur very
important). Nevertheless using this approximation we can feel the
gross-behavior of the transition probabilities coming from the
transition charge density distribution. Let us consider the simple
description of the K-electrons  conversion process starting with
the Hamiltonian \cite{Church}

\begin{equation}
\label{ham}
H=H_{nucl}+H_{elect}-%
{\displaystyle\sum\limits_{p,e}}
\frac{\alpha}{\left\vert r_{p}-r_{e}\right\vert }%
\end{equation}
\begin{equation}\label{Hprime}
H^{\prime}=-{\displaystyle\sum\limits_{p,e}}
\frac{\alpha}{\left\vert r_{p}-r_{e}\right\vert }
\end{equation}
\[
\left\langle i\left\vert H^{\prime}(L=0)\right\vert f\right\rangle
=
\]
\begin{equation}\label{sumint}
-%
{\displaystyle\sum\limits_{p,e}}
\alpha\left[
{\displaystyle\int}
d\tau_{nuc}%
{\displaystyle\int\limits_{0}^{r_{p}}}
d\tau_{e}\phi_{f}^{\ast}\psi_{f}^{\ast}\frac{1}{r_{p}}\phi_{\mathbf{i}}%
\psi_{\mathbf{i}}+%
{\displaystyle\int}
d\tau_{nuc}%
{\displaystyle\int\limits_{r_{p}}^{\mathbf{\infty}}}
d\tau_{e}\phi_{f}^{\ast}\psi_{f}^{\ast}\frac{1}{r_{e}}\phi_{\mathbf{i}}%
\psi_{\mathbf{i}}\right]
\end{equation}
Using the approximation  \cite{Church}
\begin{equation}\label{si}
\sum_{p}\int
d\tau^{\prime}\phi_{j}^{\ast}\phi_{j}\delta(r_{p}^{\prime}-r)
\end{equation}
and electron initial wave function ${\sim e^{-ar}}$ and ${\sim
e^{i\bf {kr}}}$ in infinity. In the case  of cut-off charge
density distribution  $d_0 \Theta (R - r)$ for K-electrons we find
the result of the above integration (\ref{sumint}) as
\begin{equation}\label{nucel}
F_{nuc,el}(k,R)=\frac{16 \pi ^2 \alpha  \left(k R \left(k^2
R^2+3\right) \cos (k R)-3 \sin (k R)\right)}{3 k^5}
\end{equation}
It is very important for our further consideration that this
nucleus-electron factor depends on the nucleus size $R$ and
electron impulse defined by nuclear transition energy. To make the
sense of this sentence more understandable let $d(r,R)$ be any
nuclear characteristic that depends on nuclear size R. Thus we can
write
\begin{equation}
d(r,R)=\frac{1}{2\pi}\int_{-\infty}^{\infty}\int_{-\infty}^{\infty}%
d(r,R_{0}+x) e^{ip(x-\Delta R(b^{+},b))} dpdx
\end{equation}
and following the receipt of paper  \cite{{IJMP}} we introduce the
monopole degrees of freedom and then the expectation values of
this characteristic between different collective states is
determined by matrix elements
\begin{equation}
\label{ag} <n_1\mid d(r,R)\mid n_2>
\end{equation}
\begin{equation}
=\frac{1}{2\pi}\int_{-\infty}^{\infty}\int_{-\infty}^{\infty}%
d(r,R_{0}+x)e^{ipx} <n_1\mid e^{-ip\Delta R(b^{+},b)}\mid n_2>
dpdx
\end{equation}
For illustration of this approach let take $d(r,R)=d_0
\theta(R-r)$ and calculate the matrix elements for the ground
state remembering that
\begin{equation*}
<0\mid e^{-ip\Delta R(b^{+},b)}\mid 0>= e^{\frac{-p^2S^2}{2}}
\end{equation*}
Performing integration by  $dp$ and $dx$ in (\ref{ag})we have
\begin{equation}\label{den}
\int_{-\infty}^{\infty}\int_{-\infty}^{\infty}\frac{d_0 \theta (R
- r-x)}{2\pi}e^{i p x - \frac{p^2 s^2}{2}}dpdx=\frac{1}{2}d_0
Erfc(\frac{r-R}{\sqrt{2} s})
\end{equation}

    And as a result of integration by $dpdx$ we obtain instead of
starting  cut-off function the function with surface diffuseness
(see Figure \ref{density}):

\begin{equation}
\theta(R-r)\,\,\, \Rightarrow \,\,\,\frac{1}{2}\,d_0 \,Erfc
[\frac{r-R_0}{\sqrt{2} s}]
\end{equation}
And the corresponding mean square radius is :
\begin{equation}
\left< r_{ms}^{2}\right> _{00}^{1/2}=\sqrt{\frac{%
0.6R_{0}^{5}+6R_{0}^{3}S^{2}+rR_{0}S^{4}}{R_{0}^{3}+3R_{0}S^{2}}}
\label{Rms00}
\end{equation}
It is clear that in the case of $S=0$ we have the cut-off density
mean square radius
\begin{equation}
\left< r_{ms}^{2}\right> _{00}^{1/2}=\sqrt{\frac{3}{5}}\, R_{0}
\end{equation}
\hspace{0.3in}Small vibrations of nuclear shapes around
equilibrium can give rise to physical states at low to moderate
excitation energies.

     The description of the vast amount of experimental data on
the low-lying collective spectra of even-even nuclei in the
rare-earth and actinide regions is still a problem of particular
interest in the nuclear structure physics. The classifications of
this data is mainly done from a "horizontal perspective" in
sequences of nuclei where the investigated nuclear characteristics
are empirically studied as functions of the numbers of their
valence nucleons. The experimental information on a large
sequences of states with $J^{\pi}=0^{+}$, is in some cases enough
to use a kind of a statistical approach for the  distribution.
Many well studied nuclei can be listed around the nuclear chart.
The theoretical approaches that are able to explain and correctly
describe all the data in the same nucleus in this respect are
seemingly in debt to the experiment. Let us try to less our debt
at least with the $0^{+}$ bans starting with simple Hamiltonian

\begin{equation}
\label{HamR}H=\mathbf{\alpha}R_{+}R_{-}+\mathbf{\beta}R_{0}R_{0}+
\frac{\mathbf{\beta}\Omega}{2}R_{0}%
\end{equation}
These operators was constructed with the pairs of fermion
operators $a^{\dagger}$ and $a$ of the fermions placed at
sub-shell $j$.
\begin{equation}%
\begin{tabular}
[c]{l}%
$R_{+}={\frac{1}{2}}\sum\limits_{m}(-1)^{j-m}\alpha_{jm}^{\dagger}\alpha
_{j-m}^{\dagger}\;$,\\
$ R_{-}={\frac{1}{2}}\sum\limits_{m}(-1)^{j-m}\alpha_{j-m}\alpha_{jm}\;,$\\
$ R_{0}={\frac{1}{4}}\sum\limits_{m}(\alpha_{jm}^{\dagger}\alpha_{jm}%
-\alpha_{j-m}\alpha_{j-m}^{\dagger})\;.$\\
$ \left[  R_{0},R_{\pm}\right]  =\pm R_{\pm},\ \ \ \ \ \ \ \left[  R_{+}%
,R_{-}\right]  =2R_{0}$%
\end{tabular}
\label{ROPER}%
\end{equation}

Further applying the Holstein-Primakoff transformation to the
operators \,$ R_{+}$ , $R_{-}$\, and \, $R_{0}$%

\begin{equation}
R_{-}=\sqrt{2\Omega-b^{\dagger}b}\,b \hspace{3mm}
R_{+}=b^{\dagger}\sqrt {2\Omega-b^{\dagger}b}\hspace{3mm}
R_{0}=b^{+}b-\Omega
\end{equation}
\begin{equation}
\left[  b,b^{\dagger}\right]  =1\hspace{3mm}\left[  b,b\right]
=\left[ b^{\dagger},b^{\dagger}\right]  =0
\hspace{3mm}b|0\rangle=0 \hspace
{3mm}\hspace{3mm}|n\rangle=\frac{1}{\sqrt{n!}}b^{n}|0\rangle
\end{equation}

the initial Hamiltonian (\ref{HamR}) written in terms of ideal
bosons has the form:
\begin{equation}\label{Hb}
H=Ab^{\dagger}b-Bb^{\dagger}bb^{\dagger}b.
\end{equation}
Thus the energy spectrum produced by Hamiltonian (\ref{Hb}) \ is
the parabolic function of the number of monopole ideal bosons $n$
\begin{equation}\label{En0}
E_{n}=An-Bn^{2}+C
\end{equation}

 Such a classification of large amount of experimental data in terms of
integer classification parameter recently has been done based on
phenomenological monopole part of collective Hamiltonian for
single level approach written in terms of boson creation and
annihilation operators $R_{+}$ , $R_{-}$ and $R_{0}$

    Now we can label every $K^{\pi}=0^{+}$ state by an additional
characteristic $n$ - number of monopole bosons determining it's
collective structure. The parameters $A$ and $B$ of (\ref{Hb}) are
evaluated by fitting the experimental energies of the different
$0^{+}$ states of a given nucleus to the theoretical ones applying
all possible permutations of the classification numbers $n$ and
extracting the distribution corresponding to the minimal value of
$\chi$-square \cite{gari}. The analysis based on phenomenological
collective Hamiltonian (\ref{Hb}) have shown that the experimental
energies of low lying excited $0^{+}$ states in nuclei can be
rearranged in a manner in which the energies of these states are
distributed by number of collective excitations with parabolic
distribution function $E_{n}=An-Bn^{2}+C$.

    Estimating further the E0-transition nuclear matrix elements
between different excited $0^{+}$ states
$\frac{1}{\sqrt{m!}}(b^\dagger)^{m}|0\rangle$\,\, and\,\,
$\frac{1}{\sqrt{n!}}(b^\dagger)^{n}|0\rangle$ in the same nucleus
we automatically have the transition energy values $E(n)-E(m)$.
Now we know how to proceed - first we must calculate the
transitional matrix elements.
\begin{equation}
f(m,n,p,w) \rightarrow <n \mid e^{-ip\Delta R(b^{+},b)} \mid m>
\end{equation}
\begin{equation}\label{Fmn}
f(m,n,p,w)=<n\mid e^{-ip\Delta R(b^{\dagger},b)}\mid m>\frac{1}{\sqrt{n!m!}}\langle0|b^{n}e^{-ip\Delta R(b^{+}%
,b))}(b^{\dagger})^{m}|0\rangle
\end{equation}
Using obtained in \cite{Petia} expressions
\begin{eqnarray}
\sum\limits_{l=0}^{n-1}\frac{m!}{(m-n+l)!}\binom{n}{l}(b^{\dagger
})^{m-n+l}b^{l}\;\;\;\;\;\;\;\;\;n\leqslant m\
\\
\sum\limits_{l=0}^{m-1}\frac{n!}{(n-m+l)!}\binom{m}{l}(b^{\dagger
})^{l}b^{n-m+l}\;\;\;\;\;\;\;\;\;\;\;\;n \geqslant m\
\end{eqnarray}
we can find from (\ref{Fmn})
\begin{equation}
 f(m,n,p,w)= \frac{e^{\frac{p^{2}\,{R_{0}}^{2}\,w^{2}}{2}}}{\sqrt{n!m!}}
\sum\limits_{k=0}^{\infty}(R_{0}w)^{2k+m-n}(ip)^{2k+m-n}\frac{(m+k)!}
{k!(m+k-n)!}%
\end{equation}
and summation by k gives:
\begin{equation}
 f(m,n,p,w)=
\frac{e^{\frac{p^{2}\,{R_{0}}^{2}\,w^{2}}{2}}\,{\left(  ip\right)
}^{m-n}\,{\left(  R_{0}\,w\right)  }^{m-n}\,\Gamma(1+m)\,_{1} F_{1}%
(1+m,1+m-n,- p^{2}\,{R_{0}}^{2}\,w^{2} )}{2\,\pi\,\sqrt{m!\,n!}
\,\Gamma
(1+m-n)}%
\end{equation}

    Finally the nuclear $E0$\, transitional matrix elements is nothing
but:
\begin{equation}
\label{common}
\rho_{mn}
=\frac{A_{norm}}{2\pi}\int_{-\infty}^{\infty}\int_{-\infty}^{\infty}
F_{nuc,el}(k,R_0 +x)e^{ipx}f(m,n,p,w)dpdx
\end{equation}
\section{Results and Conclusion}

\begin{equation}
 f(m,m,p,w)=\frac{e^{\frac{p^2 w^2}{2}} \Gamma (m+1) \,
   _1F_1\left(m+1;1;-p^2 w^2\right)}{2 \pi }
\end{equation}
\begin{equation}
 f(m,m-1,p,w)=\frac{i e^{\frac{p^2 w^2}{2}} p w \Gamma (m+1) \,
 _1F_1\left(m+1;2;-p^2 w^2\right)}{2 \pi }
\end{equation}
\begin{equation}
f(m,m-2,p,w)=-\frac{e^{\frac{p^2 w^2}{2}} p^2 w^2 \Gamma (m+1) \,
   _1F_1\left(m+1;3;-p^2 w^2\right)}{4 \pi }
\end{equation}
\begin{equation}
f(m,0,p,w)=\frac{e^{-\frac{1}{2} p^2 w^2} (i p)^m w^m}{2 \pi }
\end{equation}
\\
For chosen $m$ and $n$  we can perform integration by $dpdx$ in
(\ref{common}). For instance
\begin{eqnarray*}
\rho_{1 \rightarrow 0}-\frac{5}{3} k^4\pi w^9 - \frac{20}{3}
k^4\pi R_{0}^2 w^7 + \frac{16}{3} k^2\pi w^7 - \frac{10}{3} k^4\pi
R_{0}^4 w^5 \\+ 16 k^2\pi R_{0}^2 w^5 - 8\pi w^5 - \frac{4}{9}
k^4\pi R_{0}^6 w^3 + \frac{16}{3} k^2\pi R_{0}^4 w^3 \\- 16\pi
R_{0}^2 w^3 - \frac{1}{63} k^4\pi R_{0}^8 w + \frac{16}{45} k^2
\pi R_{0}^6 w - \frac{8}{3}\pi R_{0}^4 w\\
\end{eqnarray*}
\begin{eqnarray*}
\rho_{2 \rightarrow 1}= -\frac{50}{3} k^4\pi w^9 - \frac{160}{3}
k^4\pi R_{0}^2 w^7
 + \frac{128}{3} \ k^2\pi w^7 - 20 k^4\pi
R_{0}^4 w^5 \\
+ 96 k^2 \pi R_{0}^2 w^5 -48\pi w^5 - \frac{16}{9} k^4\pi R_{0}^6
w^3 + \frac{64}{3} k^2\pi R_{0}^4 \ w^3 \\ - 64\pi R_{0}^2 w^3 -
\frac{2}{63} k^4\pi R_{0}^8 w + \frac{32}{45} k^2\pi \ R_{0}^6 w
- \frac{16}{3}\pi R_{0}^4 w\\
\end{eqnarray*}
\begin{eqnarray*}
\rho_{4 \rightarrow 1}= -\frac{400}{3} k^4\pi w^9 - 320 k^4\pi
R_{0}^2 w^7 + 256 k^2\pi w^7 - 80 k^4\pi R_{0}^4 w^5 \\
+ 384k^2\pi R_{0}^2 w^5 - 192\pi w^5 - \frac{32}{9} k^4\pi R_{0}^6
w^3 + \frac{128}{3} k^2\pi \ R_{0}^4 w^3 - 128\pi R_{0}^2 w^3
\end{eqnarray*}

    The results for different $m$ and $n$ are analytical but because
of large algebra we wont present all of them here.

    For finding-out of existence of mentioned above $0^{+}$ state
with energy 0.68 MeV in $^{160}Dy$ nucleus we measure
$\beta$-spectrograms of \cite{prop} DLNP JINR for fractions Er
(two photographic plates) and Ho (one photographic plate) using
universal installation MAC-1 in ITEP \cite{egor}. At the analysis
it was found out, that in all three photographic plates to the
left of known line EIK with energy 682.3 keV below by energy on 1
keV, the peak comparable by intensity with the specified line is
confidently observed. Our attempts to carry the mentioned peak to
a conversion line or to any of  known from the literature
\cite{lit} transitions in $^{160}Dy$ nucleus have not crowned with
success. Then we proposed, that this peak is probably caused by
new transition with energy 681.3 keV, unloading the corresponding
new raised state with energy 681.3 keV to the ground state. Except
for the specified state, from experiment the states with
excitation energies 1280.0, 1456.7, 1708.2 and 1952.3 keV are
known. Considering, that from these levels transitions to the
entered by us 681.3 keV level are possible, we have undertaken
searches of such transitions. As a result such transitions with
energy  1822.5 {1822.4(3)} $I=0.24$ , between $2^{+}$ state 2503.8
keV and $0^{+}$ state 681.3 keV, and the transition from 681.3 keV
state to $2^{+}$ with energy 86.8 keV (594.5 and  $I < 0.3$ ) have
been found out. In spite of  that this facts already are powerful
enough argument in favor of existence of a state 681.3 keV in a
nucleus $^{160}Dy$, we proceed the searches of other transitions.
\\

{\bf Caption to Figures}

{\bf Figure 1.}  Density distributions before and after involving
the collective degrees of freedom.

{\bf Figure 2.}  The behavior of calculated matrix
 elements $\rho_{m \rightarrow 0}^2$ ; $m$ is the number of monopole bosons
constructed corresponding $0^{+}$ excited state.

{\bf Figure 3.} The behavior of calculated matrix
 elements $\rho_{m \rightarrow m-1}^2$ ; $m$ is the number of monopole bosons
constructed corresponding $0^{+}$ excited state

{\bf Figure 4.} The behavior of calculated matrix elements
$\rho_{m \rightarrow m-2}^2$ ; $m$ is the numbers of monopole
bosons constructed corresponding $0^{+}$ excited state.

{\bf Figure 5.} The behavior of calculated matrix
 elements $\rho_{m \rightarrow m-3}^2$ ; $m$ is the number of monopole bosons
constructed corresponding $0^{+}$ excited state.

{\bf Figure 6.} $0^{+}$ states of $^{160}Dy$ distributed in
parabola. Red line presents our calculations, red circles point
the region of predicted states and the stars are experimental
data.

{\bf Figure 7.} $0^{+}$ states of $^{158}Gd$ distributed in
parabola. Red stars present our calculations and blue squares -
experiment.

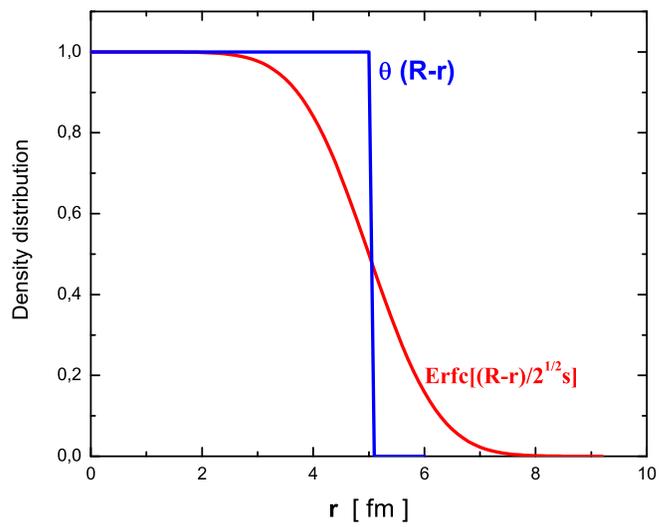
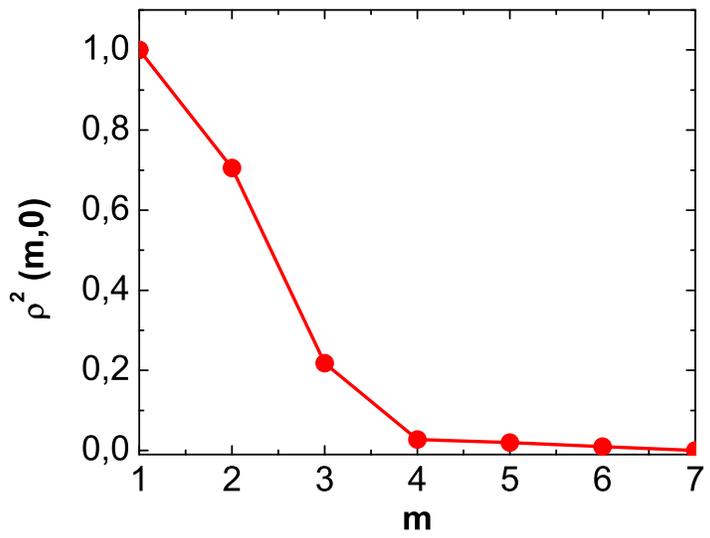
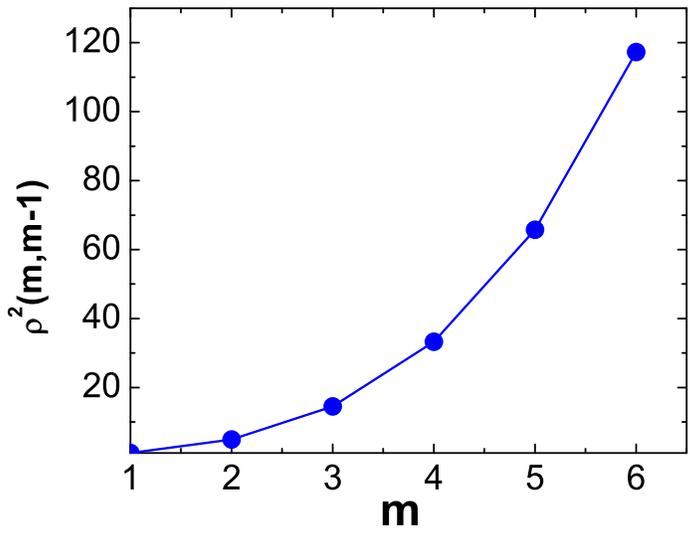
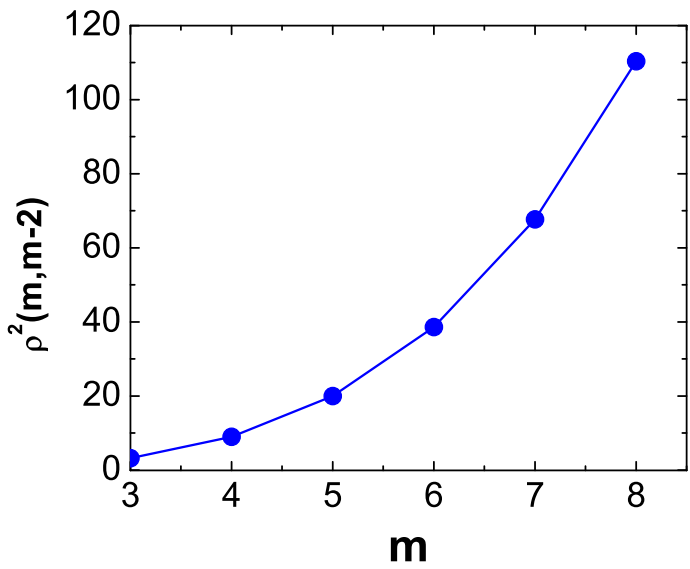
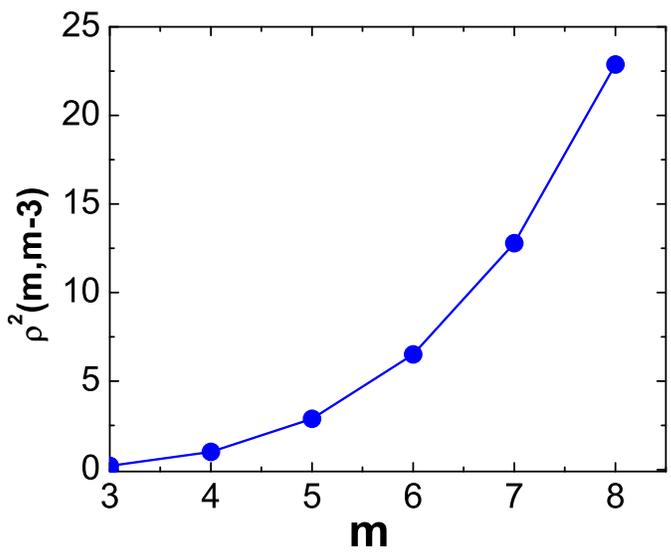
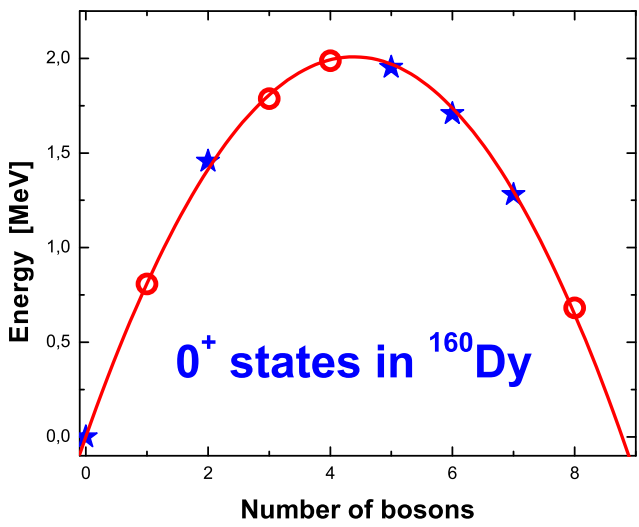
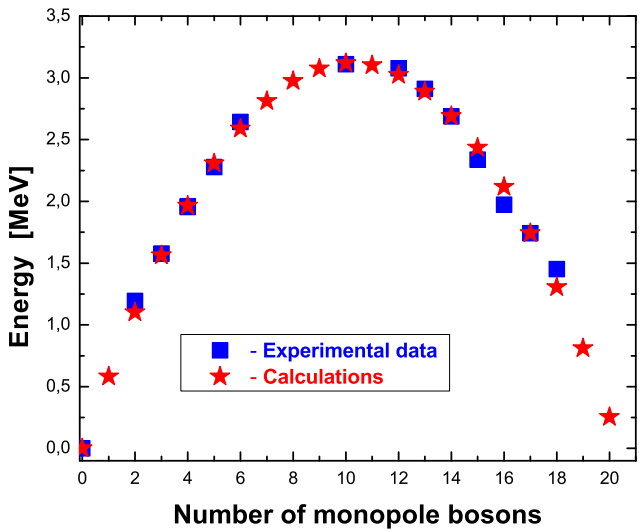
\end{document}